# Compensation of anisotropy in spin-Hall devices for neuromorphic applications

Pankaj Sethi*, Dédalo Sanz-Hernández, Florian Godel, Sachin Krishnia, Fernando Ajejas[a], Alice Mizrahi, Vincent Cros, Danijela Marković and Julie Grollier

*Unité Mixte de Physique CNRS/Thales, Université Paris-Saclay, 91767 Palaiseau, France*

Spintronic nano-oscillators with reduced non-linearity could offer key benefits for realizing neuromorphic applications such as spike-based neurons and frequency multiplexing in neural networks. Here, we experimentally demonstrate the reduction in non-linearity of a spin-Hall nano-oscillator (SHNO) by compensation of its effective magnetic anisotropy. The study involves optimization of Co/Ni multilayer growth to achieve the compensation, followed by spin diode measurements on patterned microstrips to quantify their anisotropy. The relation between the second ($H_{k2}$ = 0.47 mT) and the first order ($H_{k1}^{eff}$ = − 0.8 mT) anisotropy fields reveals the existence of an easy cone, thereby validating the presence of compensation. Furthermore, we demonstrate a synapse based on the compensated spin diode which has a fixed frequency when the input power is varied. We then study the current-induced auto-oscillation properties of SHNOs on compensated films by patterning nano-constrictions of widths 200 and 100 nm. The invariance of the resonance frequency and linewidth of the compensated SHNO with applied dc current indicates the absence of non-linearity. This independence is maintained irrespective of the applied external fields and its orientations. The compensated SHNO obtained has a linewidth of 1.1 MHz and a peak output power of up to 1 pW/MHz emulating a nano-neuron with a low linewidth and a fixed frequency.

[a] Present address: Department of Physics and Center for Advanced Nanoscience, University of California, San Diego, La Jolla, CA, 92093, USA

*Corresponding author: pankaj.sethi@cnrs-thales.fr, pankaj8684@gmail.com

# I. INTRODUCTION

Spintronic nano-oscillators with their low device footprint, rich dynamics and multifunctionality can provide an energy efficient solution to realize neuromorphic applications [1–4]. Non-linearity is prevalent in the magnetization dynamics of such nano-oscillators. In a non-linear auto-oscillator, the frequency, $f$, has a component which depends on the precession amplitude or the effective magnetization given by,

$$f = f_{FMR} + Np, \qquad (1)$$

where $f_{FMR}$ is the frequency at ferromagnetic resonance, $N$ is the non-linear frequency shift coefficient and $p$ is the term related to the amplitude of precession [5]. This non-linearity emerges from an effective anisotropy in the system, which results in non-circular trajectory of the precessing magnetization. It leads to a large frequency tunability with current which provides multifunctionality to these nano-oscillators such as the possibility to be modulated or synchronized. This has been exploited in realizing numerous applications relevant to data communication [6–10] and neuromorphic computing [3,4,11,12]. However, there are certain systems where it is possible to reduce the effective anisotropy and, as a result, the non-linearity. In such compensated systems, the anisotropy field is counterbalanced by the demagnetization field, resulting in circular trajectories of the precessing magnetization. The absence of nonlinearity, which results in a constant frequency with respect to the injected input power or current, also offers key benefits for realizing neuromorphic applications. For instance, multiply-and-accumulate (MAC) operations using spintronic resonators employ frequency multiplexing to uniquely address input radio frequency (RF) signals from neurons to the corresponding resonators [13,14]. This requires the neurons and the corresponding spin diode-based synapses to resonate at a relatively fixed frequency independent of the injected RF power, which can be accomplished by compensating anisotropy in spintronic nano-oscillators and spin diodes, respectively. Secondly, an absence of non-linearity can reduce the phase noise of the nano-oscillator by removing the effect of amplitude noise on it. The phase noise, $\Delta f$, of an auto-oscillator is given by,

$$\Delta f = \Delta f_{thermal}\,(1 + N^2/\Gamma_{eff}^2), \qquad (2)$$

where $\Delta f_{thermal}$ is the contribution from the thermal generation linewidth and $\Gamma_{eff}$ is the effective damping [5]. The second term, which is the contribution from the amplitude noise, can be neglected if $N$ is very small. Thus, a neuron with low linewidth and a relatively fixed frequency can be realized using anisotropy compensation. A third application is the realization of spike-based neurons which was recently demonstrated via macro-spin approach and micromagnetic simulations [15]. It was shown that anisotropy compensation in a spin Hall geometry results in circular trajectories of the precessing magnetization and the resulting output is a chain of spikes emulating the biological neurons. Thus, it is important to study systems with compensated anisotropy.

Recently, Jiang *et al*. have demonstrated a linewidth reduction of spin-valve based spin-torque nano-oscillators (STNOs) by controlling the perpendicular magnetic anisotropy (PMA) of their films using He-ion irradiation [16]. An alternate planar geometry based on heavy metal and ferromagnetic layers, which utilizes spin current injected from the heavy metal by spin Hall effect to sustain precession in the ferromagnet, benefits from ease of fabrication [17–19]. Moreover, spin Hall nano-oscillators (SHNOs), in the form of a nano-constriction geometry of these layers, exhibit auto-oscillations by way of mode confinement in a potential well formed by non-uniform magnetic field [20–22]. Divinskiy *et al*. demonstrated the suppression of nonlinear damping by compensation of in-plane dipolar anisotropy with PMA in Co/Ni based disks patterned on Pt heavy metal [23]. However, the detection of auto-oscillations was performed by optical methods which are less suitable for on chip applications.

Here, we experimentally demonstrate, by all-electrical measurements, a reduction of non-linearity and linewidth of an SHNO, based on Co/Ni multilayers with compensated anisotropy and a Pt heavy metal layer. Compensation is achieved by tuning the thicknesses of the Co/Ni multilayers. The effective anisotropy is estimated using spin diode measurements performed on microstrip waveguides. The relation between the second and the first order anisotropy terms indicate the presence of an easy cone state [24–26] which validates the existence of compensation. The compensated spin diode thus obtained, does not show variation of its frequency with the injected RF power and can function as a synapse. Nano-constriction based SHNOs with different widths are then patterned on the compensated stacks and the output microwave spectra are analysed. The frequency is found to remain nearly constant as a function of dc current for a wide range of magnetic field strengths and orientations. Moreover, an extremely low linewidth close to 1 MHz (quality factor = 7500) is obtained, which does not increase significantly at large applied dc currents. Control SHNO fabricated with an in-plane anisotropy $Ni_{81}Fe_{19}$/Pt stack exhibits significant shift of frequency and linewidth with the applied dc current. The compensated SHNOs can thus operates as a neuron with a fixed frequency and a low linewidth.

## II. COMPENSATION OF ANISOTROPY IN SPIN HALL DEVICES

### A. Sample preparation

The stacks consisting of Ta (5) /Pt (6) /[Co (x) /Ni (y)]$_5$ /Co (x) /Al (2) (thicknesses are in nm) are deposited on high resistivity Silicon (001) substrates (resistivity > 10000 Ω-cm) by dc-magnetron sputtering at room temperature. Ta is used as a seed layer to promote adhesion between silicon and the subsequent layer and Pt serves as the heavy metal layer. Co/Ni multilayers are chosen for their large PMA and spin polarization which can be tuned by varying layer thicknesses [27], as demonstrated previously for domain-wall based devices [28,29]. A Co/Ni multilayer repetition of five was chosen to obtain a sizeable absolute magnetization [30].

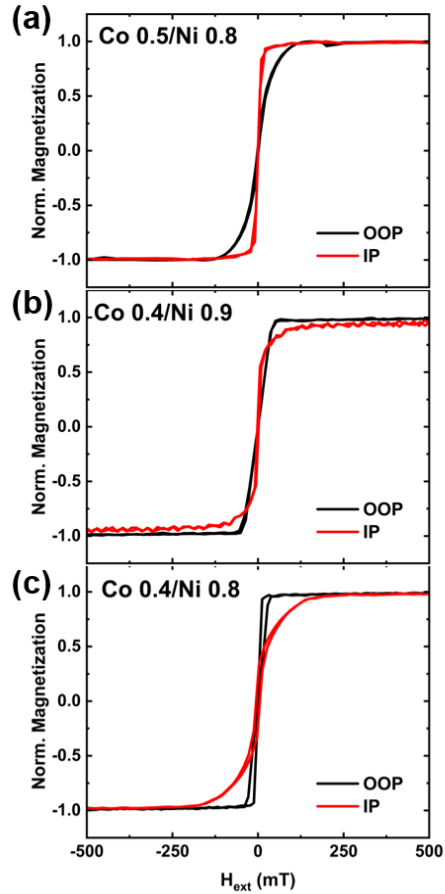

FIG. 1. Alternating gradient magnetometry measurements for Ta (5) /Pt (6)/ [Co (x) / Ni (y)]5/ Co (x)/ Al (2) (thicknesses are in nm) films with (a) in-plane anisotropy (x = 0.5, y = 0.8), (b) compensated anisotropy (x = 0.4, y = 0.9) and (c) perpendicular anisotropy (x = 0.4, y = 0.8).

Thicknesses of Co (x) and Ni (y) are varied to tune the anisotropy and the corresponding M-H loops are measured for in-plane (IP) and out-of-plane (OOP) field orientations using alternating gradient force magnetometry (AGFM). Starting with in-plane anisotropy (IPA) for Co (0.5 nm) and Ni (0.8 nm) [Fig. 1(a)], the thickness of Co is reduced to 0.4 nm and PMA is obtained [Fig. 1(c)] due to interfacial anisotropy overcoming the demagnetization field. Henceforth, in this article, Co (0.5 nm) /Ni (0.8 nm) and Co (0.4 nm) /Ni (0.8 nm) multilayers are referred to as IPA and PMA stacks, respectively. Further, when the thickness of Ni is increased to 0.9 nm, the PMA reduces but the anisotropy is neither fully in-plane nor out-of-plane [Fig. 1(b)]. As will be described in what follows, the intermediate anisotropy obtained with Co (0.4 nm) and Ni (0.9 nm) has been compensated and this film is referred to as the compensated stack. The anisotropy fields were extracted using spin diode measurements [18,31]. To carry out the measurements, the multilayers were patterned into microstrip waveguides of width 10 µm and length 25 µm using optical lithography and Ar ion beam etching techniques. Ti (15 nm)/Au (150 nm) metal stacks are deposited as electrodes and patterned into coplanar waveguides overlaying the microstrips using optical lithography and lift-off techniques. The resulting samples are henceforth referred as IPA, PMA and compensated devices, respectively.

### B. Spin-diode measurements and estimation of effective anisotropy

Figure 2 shows the spin-diode measurement set-up. A microwave current with a power of 8 mW (9 dBm) is injected into the microstrip device to generate microwave frequency spin-orbit torque (SOT) on the ferromagnetic layers due to the heavy metal Pt [18]. The mixing between the oscillating magneto-resistance and the microwave current produces a dc rectified voltage, $V_{dc}$, at the ferromagnetic resonance, which is detected by using a lock-in amplifier. The external field is swept close to the OOP direction for the PMA device ($\theta$ = 5 deg) and is swept in-plane ($\varphi$ = 45 deg) for the compensated and IPA devices. By keeping the field orientation close to the anisotropy of the devices we can eliminate the artefacts due to geometry induced local anisotropy variation and simplify the analysis [32]. All measurements are performed at room temperature. Resonance plots obtained for the PMA, the compensated and the IPA devices are shown in Figures 3 (a), (b) and (c), respectively. The amplitudes observed in the resonance plots are not corrected for the non-flat frequency response of the wire bonds and the cabling in the set-up. However, in our analysis we are only interested in the estimation of the resonance fields which are independent of amplitude losses. The plots can be well fit by the sum of symmetric and antisymmetric Lorentzian curves [18]. The resonance field, $H_r$ is extracted for each of the injected microwave frequency ($f_{res}$) and the Kittel functions ($f_{res}$ vs $H_r$) are plotted for each of the three configurations. The linear relation obtained in Figure 3 (d) for the PMA device is well explained by the Kittel formula, $f_{res} = \gamma/2\pi(H_r - \mu_0 M_{eff})$ [33], where $\mu_0 M_{eff} = \mu_0 M_s - H_k$, is the effective anisotropy field. The fit of the equation yields an $M_{eff} = -35$ mT. The negative sign of $M_{eff}$ confirms the existence of PMA. Figures 3 (e) and (f) depict the $f_{res}$ vs $H_r$ plots for the compensated and the IPA devices, respectively which are well fit with the equation, $f_{res} = \gamma/2\pi[H_r(H_r + \mu_0 M_{eff})]^{1/2}$ [18]. The extracted values of $M_{eff}$ are 0.5 mT and +86.6 mT for the compensated and the IPA devices, respectively. As a comparison, the Kittel function corresponding to $M_{eff} = 0$ is also plotted together with the as obtained fits for each of the three devices. Clearly, the compensated stack-based device is closest to the near zero effective anisotropy.

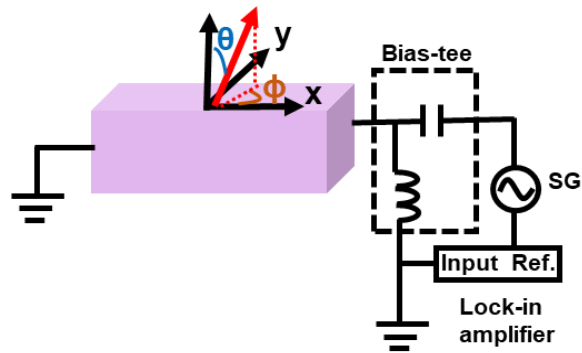

FIG. 2. Schematic illustration of spin-diode measurement set-up

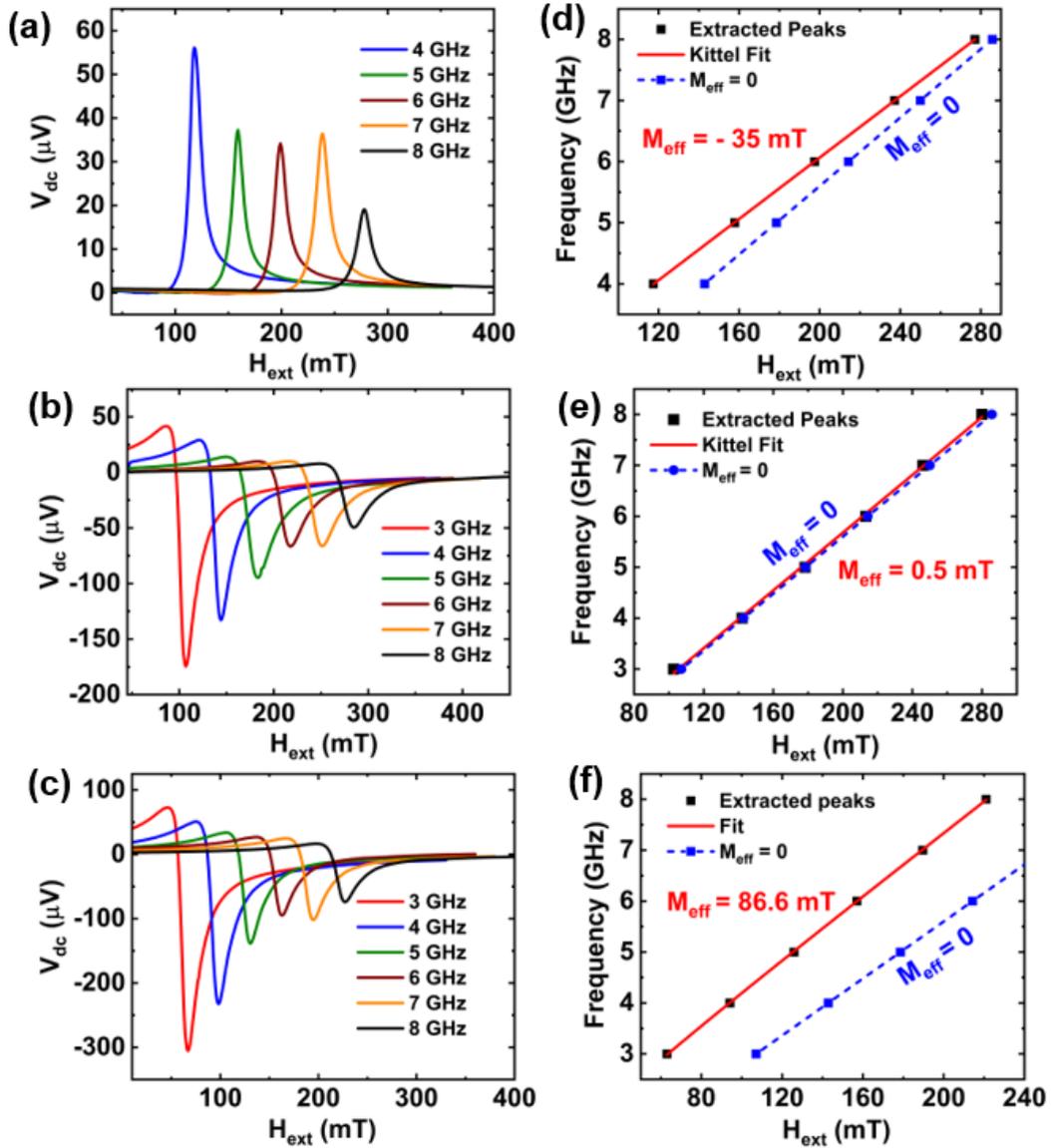

FIG. 3. Spin diode resonance plots at different injected microwave frequencies for (a) PMA, (b) compensated and (c) IPA stacks based microstrip waveguides. Resonance frequency as a function of the resonance field for (d) PMA, (e) compensated and (f) IPA stacks based microstrip waveguides. Solid red lines are Kittel fits and dotted blue lines, plotted for guidance, corresponds to $M_{eff} = 0$.

Given that the first order anisotropy is close to zero in the compensated device, the possible influence of the second order anisotropy needs to be taken into consideration. The following equations are the more generalized forms which take the second order anisotropy into consideration,

$$f = \gamma/2\pi(H_1 H_2)^{1/2} \qquad (3)$$

with

$$H_1 = H_r \cos(\theta_H - \theta_M) + H_{k1}^{eff} \cos^2\theta_M - H_{k2}\cos^4\theta_M,$$
$$H_2 = H_r \cos(\theta_H - \theta_M) + H_{k1}^{eff}\cos 2\theta_M - H_{k2}/2(\cos 2\theta_M + \cos 4\theta_M), \qquad (4)$$

where $\theta_H$, $\theta_M$ correspond to the angle of the external magnetic field and the magnetization angle measured from the sample normal, respectively. $H_{k1}^{eff}$ and $H_{k2}$ correspond to the first and the second order effective anisotropy fields, respectively [34]. By adopting $H_{k1}^{eff}$, $H_{k2}$ and $\gamma$ as adjustable parameters, the $\theta_H$ dependence of $H_r$ yields the first and the second order anisotropy fields. The energy minimum conditions $\partial F/\partial \theta_M = 0$ and $\partial^2 F/\partial \theta_M^2 > 0$ are used to extract the value for $\theta_M$, where $F$ is the magnetic energy density [34].

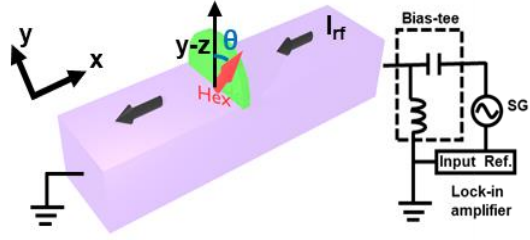

FIG. 4. Schematic illustration of spin-diode measurement set-up when external field is rotated out-of-plane.

Spin-diode measurements are performed by sweeping the magnetic field at different out-of-plane angles, $\theta_H$, in the *y-z* plane as shown in the schematic of Figure 4. In this geometry, the signal strength of the output voltage is larger due to the spin pumping contributions [35]. The resonance fields, $H_r$, are extracted from the sum of symmetric and antisymmetric Lorentzians for each of the angles. The measurements are first performed for the IPA and the PMA devices. The extracted $H_r$ as a function of $\theta_H$ are shown in Figures 5 (a) and (b), with input microwave frequencies fixed at 3 GHz and 4 GHz for the IPA and the PMA devices, respectively. The curves display a monotonic behaviour, where the $H_r$ is minimum close to the in-plane angle ($\theta_H = \pm 90$ deg) for the IPA device and close to the out-of-plane angle ($\theta_H = 0$ deg) for the PMA device. The nature of the curves is independent of the input microwave frequency, different values are selected for the two devices based on the signal quality. The measurements have been performed for the compensated device at a frequency of 5 GHz and the corresponding $H_r$ vs $\theta_H$ plots are shown in Figure 5 (c). The curves display a non-monotonic behaviour, where the $H_r$ is minimum at an intermediate angle close to 50 deg. This is referred to as the cone angle and its existence is an indication of compensation of the anisotropy [24,36]. The curves are well fit with (4) and are used to extract $H_{k1}^{eff} = -0.8$ mT and $H_{k2} = 0.47$ mT. The obtained parameters also satisfy the following conditions for the existence of an easy cone: $H_{k1}^{eff} < 0$; $H_{k2} > 0$ and $H_{k2} > -H_{k1}^{eff}/2$ [24]. These measurements thus demonstrate that a device with compensated anisotropy has been fabricated that can be employed to realize a synapse with a fixed frequency.

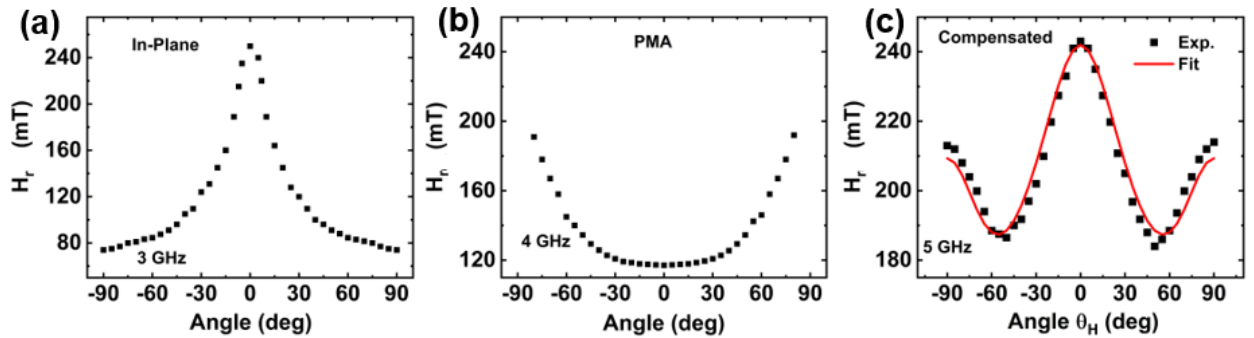

FIG. 5. Resonance field vs field angle for the microstrip waveguide with (a) IPA stack, microwave frequency fixed at 3 GHz (b) PMA stack, microwave frequency fixed at 4 GHz and (c) compensated stack, microwave frequency fixed at 5 GHz.

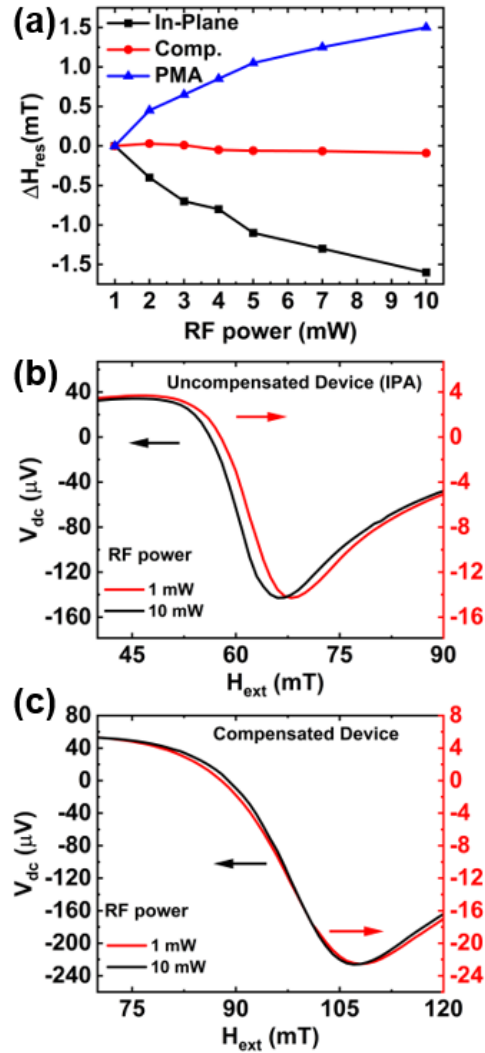

FIG. 6. (a) Comparison of shift in resonance field as a function of input rf power for a spin diode in IPA, compensated and PMA configuration. (b) Resonance curves as a function of input rf power for (b) IPA and (c) compensated (synapse) spin diodes

## C. Input independent spin-Hall synapse with fixed frequency

A synapse can be realized using spin-diodes. Leroux *et al*. demonstrated a MAC operation using magnetic tunnel junctions as spin diodes [14]. In a MAC operation, the output voltage $U_j$ can be represented by a weighted sum of the input power, $U_j = \Sigma P_i W_{ji}$. The above equation can be mapped to a spin-diode equation in the linear zone close to resonance, where the weights are represented by the resonator frequencies. During the frequency multiplexing in a MAC operation, each injected input power $P_i$, should be able to uniquely address the corresponding synapse by its frequency. This imposes a constraint on the frequency of the synapse which should not change with the injected rf power. In a spintronic resonator, this criterion is usually not satisfied on account of the inherent non-linearity. However, the compensated spin diode can be operated as an input independent synapse with a fixed frequency. Figure 6 (a) shows the shift in $H_r$ as a function of the injected input rf power for the IPA, the compensated and the PMA spin diodes. Starting at the minimum input power ( = 1 mW), the shift is normalized to 0 for all the three devices. As the input power is increased, the IPA and the PMA devices exhibit an increase in the shift of $H_r$, whereas, the compensated device shows a negligible shift in $H_r$. Figure 6 (b) and (c) show the comparison of the resonance plots for the IPA and the compensated devices, respectively, as a function of the input power. Clearly, there is no visible shift in the resonance field and the equivalent frequency with the injected rf power for the compensated device as compared to the IPA device. Thus, the compensated spin diode can function as an input independent spin-Hall synapse.

## III. AUTO-OSCILLATIONS IN COMPENSATED SPIN HALL DEVICES – NEURON OPERATION

### A. Device fabrication and measurement set-up

Nano-constrictions with widths of 100 nm and 200 nm are fabricated on the compensated Co/Ni stacks using electron-beam lithography and Ar ion beam etching. Ti (15 nm)/Au (150 nm) metal stacks are deposited as electrodes and patterned into coplanar waveguides overlaying the nano-constrictions using optical lithography and lift-off. The device geometry is similar to the one used in previous reports for realizing an SHNO [21,22]. As a comparison, in-plane SHNO based on Py/Pt stacks are also patterned into nano-constrictions (Py = Permalloy = $Ni_{81}Fe_{19}$).

The scanning electron microscopy image of a 200 nm nano-constriction along with the measurement set-up to detect the auto-oscillations is shown in Figure 7 (a). A dc current, $I_{dc}$, is injected into the nano-constriction via the dc port of a bias-tee. An external magnetic field is applied at an in-plane angle, $\phi_H$ and an out-of-plane angle, $\theta_H$. The SHNO emits microwave power which is extracted from the rf port of the bias-tee and amplified by 38 dB using a low noise wide-band amplifier. The output spectra are sampled using a spectrum analyzer. All measurements are performed at room temperature.

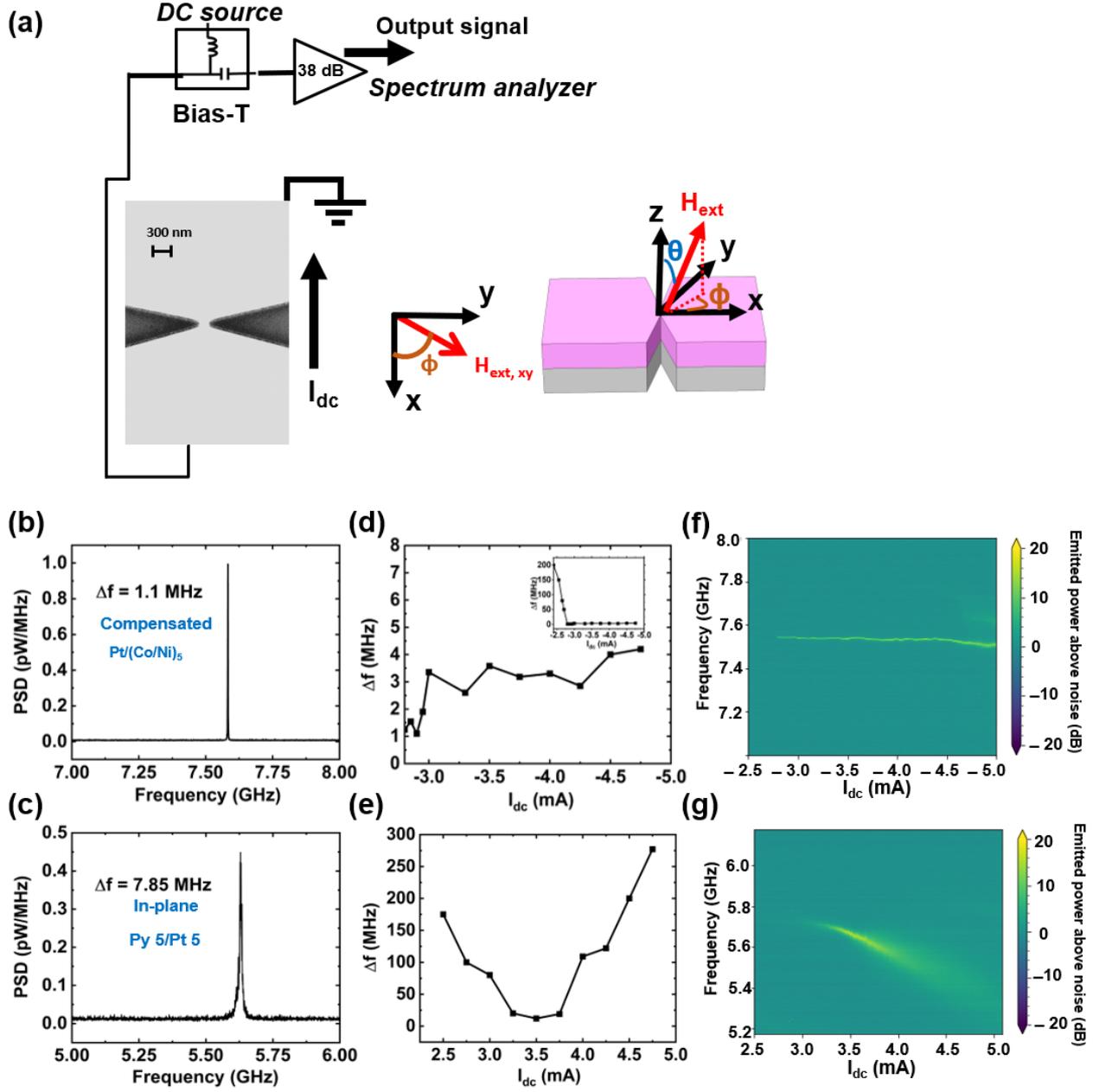

FIG. 7. (a) SEM image of 200 nm nano-constriction and a schematic to study the microwave emission from the SHNO. (b) Auto-oscillation spectra for the compensated Co/Ni SHNO obtained at $I_{dc} = +2.8$ mA, $H_{ext} = 300$ mT ($\theta_H = 15$ deg, $\phi_H = 50$ deg). (c) Auto-oscillation spectra for the in-plane Py/Pt SHNO obtained at $I_{dc} = -3.5$ mA, $H_{ext} = 50$ mT ($\theta_H = 85$ deg, $\phi_H = 42$ deg). Linewidth as a function of $I_{dc}$ sweep for (d) compensated Co/Ni SHNO and (e) in-plane Py/Pt SHNO. Power spectral density plots showing frequency vs $I_{dc}$ sweep for (f) compensated Co/Ni SHNO and (g) in-plane Py/Pt SHNO.

**B. Electrical microwave measurements for compensated and in-plane devices**

Figure 7 (b) shows the emission spectra for the 200 nm SHNO realized using the compensated Co/Ni stack at $I_{dc}$ = −2.8 mA (+ x-direction) and $H_{ext}$ = 300 mT ($\theta_H$ = 15 deg, $\phi_H$ = 50 deg). The linewidth ($\Delta f$) obtained from the Lorentz fit is 1.1 MHz with the peak power spectral density (PSD), after subtracting the amplifier gain, as high as 1 pW/MHz. To the best of our knowledge, the quality factor (Q ≈ 7500) obtained is more than the highest reported using a single constriction based SHNO [3,37]. As a comparison, the above measurements are also performed on Py/Pt based SHNO devices. Figure 7 (c) shows the corresponding spectra obtained at $I_{dc}$ = +3.5 mA and $H_{ext}$ = 50 mT ($\theta_H$ = 85 deg, $\phi_H$ = 42 deg). It is worth noting that the field orientation is maintained close to the in-plane direction for this device to excite the in-plane modes and the sign of $I_{dc}$ is positive as the SOT is from the top interface. The minimum linewidth obtained from the Lorentz fit is 7.85 MHz and is much larger than that achieved using the compensated Co/Ni SHNO. The above observations can be explained from (2), which indicate a reduction of $\Delta f$ if N reduces. To further validate this claim, we sweep the injected $I_{dc}$ and record the variation of the frequency and $\Delta f$ for the two SHNOs at the above-mentioned external fields and orientations, respectively. Figures 7 (d) and (e) show $\Delta f$ as a function of $I_{dc}$ for the compensated Co/Ni and the in-plane Py/Pt SHNOs, respectively. Figure 7 (d) is plotted for $I_{dc}$ larger than the critical current of auto-oscillations ($I_c$ = − 2.7 mA), which is the region of interest, and the inset shows the data for $I < I_c$ as well. When $I_{dc} < I_c$, $\Delta f$ increases with the reduction in current for both the devices, as expected. At large $I_{dc}$, the Py/Pt SHNO shows an increase in $\Delta f$ due to the inherent non-linearity, which is not the case with the compensated Co/Ni SHNO which shows a near constant $\Delta f$. The evidence for the absence of non-linearity in the compensated Co/Ni SHNO becomes stronger when we compare its frequency vs $I_{dc}$ shown in the power spectral density plots in Figure 7 (f) to that obtained for Py/Pt SHNO in Figure 7 (g). Clearly, the rate of change of frequency with the current (*df/dI*) is minimal for the compensated Co/Ni SHNO (= 10 MHz/ mA) and significant for the in-plane Py/Pt SHNO (= 500 MHz/mA). However, for $I_{dc} > −4.5$ mA, some non-linearity can be observed in Figure 7

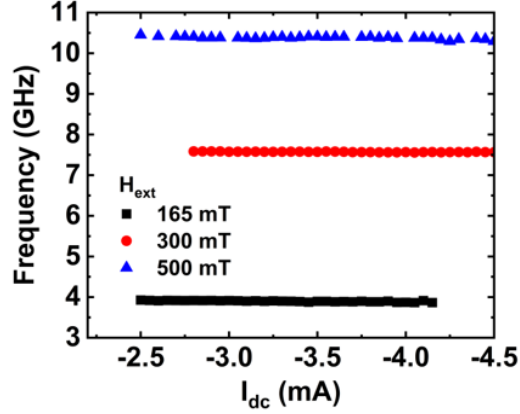

FIG. 8. Auto-oscillation frequency as a function of $I_{dc}$ sweep for compensated Co/Ni SHNO performed at external fields of 165 mT, 300 mT and 500 mT.

(f), which could be ascribed to the device heating or frequency shift due to the Oersted field or the field-like torque [30]. The above observations are a direct validation of a reduction in the non-linearity as indicated in (1). The measurements are repeated at different applied external magnetic fields to the compensated Co/Ni SHNO and are shown in Figure 8. As is the case, the external fields only change the frequency of the ferromagnetic resonance and not the slope which are nearly zero for the compensated Co/Ni SHNO.

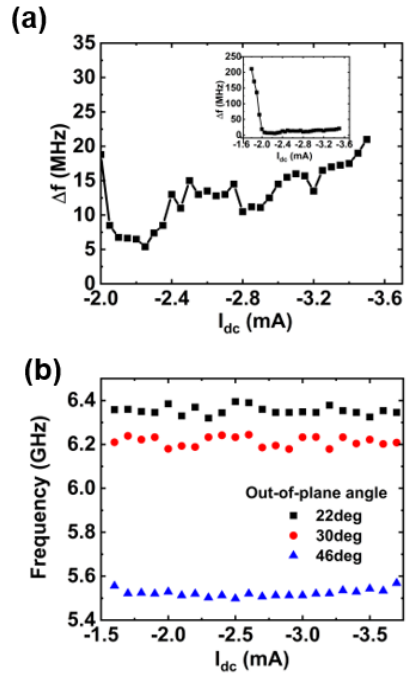

FIG. 9. (a) Linewidth as a function of $I_{dc}$ sweep at $H_{ext}$ = 180 mT ($\theta_H$ = 15 deg, $\phi_H$ = 50 deg) for the compensated Co/Ni SHNO with 100 nm width. (b) Comparison of frequency vs $I_{dc}$ sweep when $H_{ext}$ = 180 mT is applied along out-of-plane angles of 22, 30 and 46 deg to the 100 nm compensated Co/Ni SHNO

To further validate the existence of compensation across different devices, the measurements are repeated on a 100 nm constriction. Figure 9 (a) shows the variation of $\Delta f$ vs $I_{dc}$ for this device, performed at $H_{ext}$ = 180 mT ($\theta_H$ = 15 deg, $\phi_H$ = 50 deg). The plot indicates a high $\Delta f$ for $I_{dc} < I_c$ (= −1.8 mA), as shown in the inset, upon which it does not increase significantly at higher currents. A larger $\Delta f$ in excess of 5 MHz as opposed to 1.1 MHz is obtained when the width of the constriction is reduced from 200 to 100 nm, which is expected due to a smaller mode volume. We also performed frequency vs $I_{dc}$ for this device at different orientations of the external magnetic field ($H_{ext}$ = 180 mT). The measurements are performed for three different angles, $\theta_H$ = 22, 30 and 46 degrees, respectively keeping $\phi_H$ fixed at 90 deg. Figure 9 (b) shows the results of frequency vs $I_{dc}$ at different out-of-plane angles of the external field. At each angle, the frequency is different as expected, and is minimum at 46 deg which is close to the cone angle of precession. However, the frequency remains nearly constant with respect to $I_{dc}$, even at different angles, thus providing a strong evidence for the absence of non-linearity in the compensated SHNO device.

## IV. CONCLUSION

In summary, we experimentally demonstrate a strong reduction of non-linearity in the magnetization dynamics of an SHNO by compensation of its effective magnetic anisotropy. Co/Ni multilayers with a Pt heavy metal form the system for the study. The thicknesses of Co and Ni are tuned to change the magnetization anisotropy, which is estimated using spin-diode measurements. An easy cone anisotropy is obtained for the compensated stack when the PMA is counterbalanced by the demagnetization field. The relation between the second and the first order anisotropy fields thus obtained, satisfies the condition for the existence of an easy cone. The spin-diode signal is shown to be independent of the input power as required to operate as a synapse in neuromorphic computing applications. Auto-oscillations in the SHNO are examined using nano-constrictions fabricated from the compensated stacks and are compared with the emission spectra of Py/Pt based SHNO with an in-plane anisotropy. The frequency and the linewidth are found to be independent of the applied dc current for the compensated SHNO even at different external fields and orientations. The linewidth obtained is as low as 1.1 MHz and the peak emission power is as high as 1 pW/MHz. Thus, the compensated SHNO can operate as an artificial neuron with a fixed frequency and a low linewidth. This study opens up a possibility of realizing neuromorphic applications such as frequency multiplexing in a multiply-and-accumulate (MAC) operation, and spike-based neurons exploiting easy-plane oscillations in a compensated SHNO.

## ACKNOWLEDGMENTS

This work is supported by the Agence Nationale de la Recherche Project ANR-20-CE24-0002 (SpinSpike). J.G. and D. H. S. acknowledge support from Q-MEEN-C, an Energy Frontier Research Center funded by the U. S. Department of Energy, Office of Science, Basic Energy Science, under Grant No. DE-SC0019273, for work on neuromorphic computing with SHNO.